\documentclass[11pt]{article}
\usepackage[latin1]{inputenc}
\usepackage[english]{babel}
\usepackage[namelimits]{amsmath}
\usepackage{authblk}
\usepackage{amssymb}
\usepackage{amsmath}
\usepackage{amsthm}

{\tiny}

\begin{document}
\title{{\bf On the perturbative formalism and a  possible quantum discrete spectrum for the Regge-Wheeler equation of a de Sitter spacetime}}
\author[1,2,3]{S. Viaggiu\thanks{s.viaggiu@unimarconi.it and viaggiu@axp.mat.uniroma2.it}}
\affil[1]{Dipartimento di Scienze Ingegneristiche, Universit\'a degli Studi Guglielmo Marconi, Via Plinio 44, I-00193 Roma, Italy.}
\affil[2]{INFN, Sezione di Roma 3, I-00146 Rome Italy.}
\affil[3]{Dipartimento di Matematica, Universit\`a di Roma ``Tor Vergata'', Via della Ricerca Scientifica, 1, I-00133 Roma, Italy.}
\date{\today}\maketitle

\begin{abstract}
In this paper we study the perturbative regime in the static patch of de Sitter metric in the Regge-Wheeler formalism. After realizing that perturbative regime in a de Sitter spacetime 
depicted in terms of usual spherical coordinates cannot be extended up to the cosmological horizon, we study  
perturbative equations, in particular the axial ones, in terms of the tortoise coordinate $r_*$. We show that perturbative regime can be extended up to the cosmological horizon, provided that suitable boundary conditions are chosen. As an application, we explore the 
Regge-Wheeler equation at short distances by performing a taylor expansion. In order to study some possible quantum effects at short distances, we impose to the equation so obtained the same boundary conditions suitable for a quantum 3D harmonic oscillator. As a result, a discrete spectrum can be obtained. The aforementioned spectrum is analysed and a
relation with possible effects denoting quantum behavior of gravitons is suggested.
\end{abstract}

\section{Introduction}

The detection of gravitational waves (GW) \cite{1} (GW150914 event) represented the birth of the gravitational wave astronomy and a fundamental confirmation of General Relativity. In this regard, de Sitter universe represents a very useful arena to test the behavior of GW in a non asymptotically flat spacetime (see for example \cite{2,3,4,5} and  \cite{5a,5b} for a study in terms of Laplace transform). Generally, a perturbation of a spacetime with suitable boundary conditions at spatial infinity or at the cosmological horizon generates  quasinormal modes (see for example \cite{6,7,8,9}) with a discrete spectrum for frequencies and with a complex part describing damping modes. Normal modes are the ones with real frequencies. In \cite{R1} and references therein  has been shown that 
normal modes in 4-dimensions are not allowed in de Sitter spacetime.
Quasinormal modes are also used to study primordial quantum fluctuations and are often calculated by using 
a WKB approximation method developed in \cite{10,11} (see also \cite{12} and references therein).
However, it is possible to explore different boundary conditions with respect to the ones leading to quasinormal modes, As an example, in \cite{R2,R3} a purely real discrete spectrum of trapped gravitons in a spherical box is obtained by imposing Dirichlet boundary conditions. In particular, the vanishing
of the Regge-Wheeler function is imposed on the boundary of the spherical box and the quantum nature of the so obtained spectrum is studied in \cite{R2,R3}.\\
The aim of this paper is twofold. In the first part we study the viability of the perturbative formalism 
in de Sitter case up to the cosmological horizon. In fact,
perturbations are often performed by means of the technique developed in \cite{13,14} and further 
analysed in \cite{15,16,17,18,19,20} in terms of a basis of tensorial spherical harmonics. Another useful technique to study perturbations is provided by the Penrose formalism (see \cite{15} and references therein) in terms of optical scalars.\\
In \cite{21} it has been shown that, by using the perturbative method in \cite{13,14,15} with usual 
spherical coordinates, perturbative regime in the static patch of a de Sitter universe cannot be extended up the cosmological horizon. The authors in \cite{21} show that approaching the horizon
some functions depicting the perturbation generated by a GW become 
unbounded, this representing a breackdown of the perturbative regime.\\
Inspired by the results in \cite{21}, we reconsider perturbations in a de Sitter spacetime. 
In particular, the axial perturbations are derived and solved in the Chandrasekhar (dubbed diagonal) gauge \cite{15} in usual spherical coordinates. 
Hence, it is provided the range of validity of the perturbative regime in these coordinates. 
It is thus shown that, by using tortoise coordinate $r_*$ from the onset, perturbations can be extended
up to the horizon, provided that suitable boundary conditions 
are used.\\ 
It should be noted that in the literature the study of the range of validity of perturbative regime is
often neglected, in particular for the de Sitter case.\\
In the second part of this paper we explore a new possible discrete spectrum with a possible link with 
quantum gravity. As an example, quantum fluctuations are expected to arise on small scales with
respect to cosmological ones. Hence, 
we explore the behavior of the Regge-Wheeler equation at short distances by a taylor 
expansion 
of the Regge-Wheeler potential $V^{(a)}$, with the principal part leading to a Schrodinger-like equation with a 3D harmonic oscillator potential.\\ 
In section 2 we introduce the suitable mathematical machinery, while in section 3 we explore the range of validity of the perturbative regime. In section 4 we write down the perturbative equations in terms of
tortoise coordinate $r_*$, while, as an application, section 5 is devoted to the study of a discrete spectrum obtained by imposing suitable boundary conditions at short distances. Finally, in section 6 we outline some conclusions and final remarks.

\section{Perturbation equations}

The static patch of de Sitter metric expressed in spherical coordinates is given by:
\begin{eqnarray}
& & ds^2=e^{2\nu}dt^2-e^{-2\nu}dr^2-r^2\left(\sin^2\theta\;d{\phi}^2+d{\theta}^2\right),\nonumber\\
& & e^{2\nu}=1-H^2r^2,\;\;H=\sqrt{\frac{\Lambda}{3}}\label{1}.
\end{eqnarray}
We denote with $g_{ik}^{(0)}$ the unperturbed metric and with 
$h_{ik}$ a small perturbation $|h_{ik}|<<|g_{ik}^{(0)}|$:
\begin{equation}
g_{ik}=g_{ik}^{(0)}+h_{ik}.
\label{2}
\end{equation}
The spherical symmetry of $g_{ik}^{(0)}$ given by (\ref{1}) allows to write down $h_{ik}$ in a basis of spherical tensorial harmonics \cite{13,14} thanks to the spherical Legendre polynomials 
$Y_{\ell m}(\theta,\phi), \ell\in\mathbb{N}, m\in\mathbb{Z}, m\in [-\ell, +\ell]$. The polar perturbations 
$h_{ik}^{p}$  have even parity
under parity operator: ${(-1)}^{\ell}$. Conversely,  the axial perturbations
$h_{ik}^{a}$ have odd parity:  ${(-1)}^{\ell+1}$.
In the following, we adopt the diagonal gauge present in \cite{15}-\cite{20}.
For $h_{ik}^a$ we have:
\begin{equation}
h_{ik}^a=
\begin{pmatrix}
(t) & (\phi) & (r) & (\theta)\\
0 & h_0\sin\theta\;Y_{\ell m,\theta} & 0 & -h_0\frac{1}{\sin\theta}Y_{\ell m,\phi}\\
h_0\sin\theta\;Y_{\ell m,\theta} & 0 & h_1\sin\theta\;Y_{\ell m,\theta} & 
0\\
0 & h_1\sin\theta\;Y_{\ell m,\theta} & 0 & -h_1\frac{1}{\sin\theta}\;Y_{\ell m,\phi}\\
-h_0\frac{1}{\sin\theta}Y_{\ell m,\phi} & 0 & -h_1\frac{1}{\sin\theta}\;Y_{\ell m,\phi} &
0
\end{pmatrix},
\label{3}
\end{equation}
while for $h_{ik}^p$ we have:
\begin{equation}
h_{ik}^p=
\begin{pmatrix}
(t) & (\phi) & (r) & (\theta)\\
2Ne^{2\nu} Y_{\ell m} & 0 & 0 & 0\\
0 & -2r^2\sin^2\theta\;H_{11} & 0 & -r^2 V X_{\ell m}\\
0 & 0 & -2 e^{-2\nu}L Y_{\ell m} & 0\\
0 & -r^2 V X_{\ell m} & 0 & -2 r^2  H_{33}
\end{pmatrix}. 
\label{4}
\end{equation}
The axial perturbations (\ref{3}) depend on two functions
$h_0(t,r),h_1(t,r)$, 
while the polar perturbations (\ref{4}) depend on four functions $N(t,r),L(t,r),T(t,r),V(t,r)$. Moreover
\begin{eqnarray}
X_{\ell m}(\theta,\phi) &=& 2Y_{\ell m,\theta,\phi}-2\cot\theta\;Y_{\ell m,\phi}\label{5}\\
W_{\ell m}(\theta,\phi) &=& Y_{\ell m,\theta,\theta}-\cot\theta\;Y_{\ell m,\theta}-
\frac{1}{\sin^2\theta}\;Y_{\ell m,\phi,\phi}\nonumber\\
H_{11}(t,r,\theta,\phi) &=&TY_{\ell m}+\frac{V}{\sin^2\theta}\;Y_{\ell m,\phi,\phi}+
V\cot\theta\;Y_{\ell m,\theta}\nonumber\\
H_{33}(t,r,\theta,\phi) &=&TY_{\ell m}+VY_{\ell m,\theta,\theta}.
\end{eqnarray} 
As usual, thanks to the static nature of the background metric (\ref{1}),
we can Fourier transform the metric functions depicting axial and polar perturbation with respect to general modes of frequency $\{\omega\}$. 
For the complete metric (\ref{2}) we thus obtain:
\begin{eqnarray}
ds^2 &=& e^{2\nu}dt^2-e^{-2\nu}dr^2-r^2\left(\sin^2\theta\;d{\phi}^2+d{\theta}^2\right)+\label{6}\\
&+&\sum_{\ell m}\int_{-\infty}^{+\infty}e^{\imath\omega t}d\omega
\Bigl\{2N(\omega, r)e^{2\nu} Y_{\ell m}dt^2-
2e^{-2\nu}L(\omega,r) Y_{\ell m}dr^2-\nonumber\\
&-& 2r^2\sin^2\theta H_{11}(\omega, r,\theta,\phi)d{\phi}^2-2r^2 
H_{33}(\omega, r,\theta,\phi)d{\theta}^2+\nonumber\\
&+& 2h_0(\omega,r)\sin\theta\;Y_{\ell m,\theta}\;dt d\phi-
2h_0(\omega,r)\frac{Y_{\ell m,\phi}}{\sin\theta}dt d\theta+\nonumber\\
&+& 2h_1(\omega,r)\sin\theta\;Y_{\ell m,\theta}\;dr d\phi-
2h_1(\omega,r)\frac{Y_{\ell m,\phi}}{\sin\theta}dr d\theta-\nonumber\\
&-& \left[4r^2 V(\omega,r) Y_{\ell m,\theta,\phi}-4r^2 V(\omega,r)\cot\theta\;Y_{\ell m,\phi}\right]d\theta d\phi\Bigr\}.\nonumber
\end{eqnarray}
In this paper we consider only perturbations associated to GW, i.e. $\ell \geq 2$.
In order to simplify the perturbed equations one can use a tetradic basis of four vectors $e_{(a)i}$ ($a=t,r,\theta,\phi $),
with   $e_{(a)}^i e_{(b)i}={\eta}_{(a)(b)}$ and ${\eta}_{(a)(b)}=diag(1,-1,-1,-1)$.
The field equations  are proiected onto $e_{(a)i}$, i.e. $G_{(a)(b)}=2T_{(a)(b)}+\Lambda{\eta}_{(a)(b)}$, and perturbed:
$\delta G_{(a)(b)}=2\delta T_{(a)(b)}+\delta(\Lambda{\eta}_{(a)(b)})$. For the tetrad of (\ref{1}) we have
\begin{eqnarray}
& & e_{t}^i=(e^{-\nu},0,0,0), \label{7}\\
& & e_{\phi}^i=\left(0,\frac{1}{r\sin\theta},0,0\right), \nonumber\\
& & e_{r}^i=\left(0,0,e^{\nu},0\right), \nonumber\\
& & e_{\theta}^i=\left(0,0,0,\frac{1}{r}\right), \nonumber
\end{eqnarray}
As a result, for the per perturbed equations we have:
\begin{eqnarray}
& &\delta G_{(a)(b)}=\delta\left(e_{(a)}^i e_{(b)}^k G_{ik}\right)=2\delta\left(e_{(a)}^i e_{(b)}^k T_{ik}\right),\label{8}\\
& &\delta G_{ik}=\delta R_{ik}-\frac{h_{ik}}{2}R+\frac{1}{2}g_{ik}^{(0)}R_{lm}h^{lm}-
\frac{1}{2}g_{ik}^{(0)}g^{(0)lm}\delta R_{lm}\nonumber,
\end{eqnarray}
with the perturbed tetrad given by:
\begin{eqnarray}
& & \delta e_{(t)}^i=\left[-e^{-\nu}N Y_{\ell m},e^{-\nu}\frac{h_0}{r^2\sin\theta}Y_{\ell m,\theta},0,-e^{-\nu}\frac{h_0}{r^2\sin\theta}
Y_{\ell m,\phi}\right],\nonumber\\
& & \delta e_{(\phi)}^i=\left[0,-\frac{H_{11}}{r\sin\theta},0,0\right],\nonumber\\
& & \delta e_{(r)}^i=\left[0,e^{\nu}\frac{h_1}{r^2\sin\theta}Y_{\ell m,\theta},-\frac{L}{e^{-\nu}}Y_{\ell m},
-e^{\nu}\frac{h_1}{r^2\sin\theta}Y_{\ell m,\phi}\right],\nonumber\\
& & \delta e_{(\theta)}^i=\left[0,-\frac{V}{r\sin^2\theta}X_{\ell m},0,-\frac{H_{33}}{r}\right]. \label{9}
\end{eqnarray}
For the energy momentum tensor $T_{ik}$ we obviously have $\delta T_{ik}=0$. 
Since we are considering gravitational waves traveling in the static patch of a de Sitter universe, the generation of a gravitational wave is not a consequence of the perturbation of $\Lambda$, but the gravitational wave perturbates the spacetime metric:
\begin{equation}
\delta G_{(a)(b)}= \delta(\Lambda\;{\eta}_{(a)(b)})=\Lambda\;\delta{\eta}_{(a)(b)}=0.
\label{10}
\end{equation}
In (\ref{10}) we used the fact that the tetrad ${\eta}_{(a)(b)}$ is a constant metric tensor. However, 
also note that, 
thanks to covariance, polar equations do imply $\delta\Lambda = 0$. 
In the following we focus our attention only to the axial equations. After posing $2n=(\ell-1)(\ell+2)$, the relevant equations for axial perturbations for the mode with $\{\omega,\ell,m\}$ are thus given by:
\begin{eqnarray}
& & e^{2\nu}h_{0,r,r}-\frac{2}{r^2}h_0\left(n+e^{2\nu}\right)-\imath\omega e^{2\nu}
\left(h_{1,r}+\frac{2}{r}h_1\right)=0,\label{11}\\
& & \imath\omega e^{-2\nu}h_{0,r}+\omega^2 e^{-2\nu}h_1-
\frac{2\imath\omega}{r}e^{-2\nu}h_0-\frac{2n}{r^2}h_1=0,\label{12}\\
& & \imath\omega e^{-2\nu}h_0-e^{2\nu}h_{1,r}-2\nu_{,r} e^{2\nu}h_1=0.\label{13}
\end{eqnarray}
By introducing, as usual, the Regge-Wheeler function \cite{13} $Z^{(a)}$ for the mode $\{\omega,\ell,m\}$ 
with
\begin{equation}
h_1=re^{-2\nu}Z^{(a)},\label{14}
\end{equation}
equation (\ref{13}) becomes:
\begin{equation}
\imath\omega h_0=e^{2\nu}{\left[r Z^{(a)}\right]}_{,r}. \label{15}
\end{equation}
As customary, the equation for $Z^{(a)}$ is obtained introducing the tortoise coordinate
$r_*$ by
\begin{equation}
r_*=\int_0^r e^{-2\nu} dr. \label{16}
\end{equation}
Using equation (\ref{1}) we have:
\begin{equation}
Z^{(a)}_{,r_*,r_*}+Z^{(a)}\left[\omega^2-
\frac{\left(1-H^2r^2\right)}{r^2}\ell(\ell+1)\right]=0. \label{17}
\end{equation}
After solving the (\ref{17}) for $Z^{(a)}$, from (\ref{14})
and (\ref{15}) we can obtain $h_1,h_0$

\section{Range of validity for perturbative regime in usual coordinates}

The Regge-Wheeler equation for axial perturbations is often considered in order to study quasinormal modes that are obtainded by imposing suitable boundary conditions on $Z^{(a)}$. However, it should be taken in mind that perturbative regime must hold for all functions depicting axial (and also polar) 
peturbations, as shown in \cite{21}. To be more quantitative, one introduces a small adimensional parameter $\epsilon$ representing the amplitude of the perturbation. This allows to formally depict
the perturbation by means of a series expansion  with respect to $\eta$:
\begin{equation}
g_{ik}=g_{ik}^{(0)}+
\epsilon\delta g_{ik}^{(1)}+\epsilon^2\delta g_{ik}^{(2)}+\cdots+\epsilon^n\delta g_{ik}^{(n)}
+\cdots. \label{18}
\end{equation}
As an example we have 
$\epsilon\delta g_{t\phi}^{(1)}=e^{\imath\omega t}h_0(\omega,r)\sin\theta\;Y_{\ell m,\theta}$. 
The linear regime is the one with linear terms with respect to a small parameter
$\epsilon$ . As shown in \cite{21},
with the background metric (\ref{1}), the perturbative regime cannot be extended up to the horizon
at $r=\frac{1}{H}$.\\
For our purposes, after reintroducing the speed of light $c$, equation (\ref{17}) becomes:
\begin{equation}
Z^{(a)}_{,r_*,r_*}+Z^{(a)}\left[\frac{\omega^2}{c^2}+H^2\ell(\ell+1)-
\frac{1}{r^2}\ell(\ell+1)\right]=0. \label{19}
\end{equation}
From (\ref{16}) we have:
\begin{equation}
r_*=\frac{1}{2H}\ln\left(\frac{1+Hr}{1-Hr}\right),\;r\in[0,\frac{1}{H}),\;\;
r=\frac{1}{H}\frac{\left(e^{2Hr_*}-1\right)}{\left(e^{2Hr_*}+1\right)},\;
r_*\in[0,\infty).
\label{20}
\end{equation}
With the help of (\ref{20}) we can express (\ref{19}) in terms of the tortoise coordinate or in terms of the radial coordinate $r$.\\ 
To start with, we analyse the regime with small values of $r$ (or $r_*$). Regularity condition at the origin imposes that $\lim_{r_*\rightarrow 0}Z^{(a)}(r_*)=0$ with a behavior (see below)
$Z^{(a)}\sim r_*^{\ell(\ell+1)}, \ell\geq 2$. From an ispection of (\ref{14}) and (\ref{15}) we 
conclude that both $h_1$ and $h_0$ remain in the perturbative regime.\\
From (\ref{20}) we see that in the limit for $rH<<1$, $r_*$ can be expressed in a series expansion with 
$r_*\sim r$ in the lowest order. In the regime $rH<<1$ $Z^{(a)}$ can thus be expanded in a series expansion, where at the lowest order $Z^{(a)}\sim H^{\ell(\ell+1)} r_*^{\ell(\ell+1)}$, with similar reasonings for
$h_1,h_0$. Hence, as far as $rH$ remains sufficiently 'small', perturbative regime is again valid.\\
The situation changes drastically approaching the horizon $r\rightarrow\frac{1}{H}$ or
$r_*\rightarrow\infty$. In such a limit, the asymptotic solution for $Z^{(a)}$ is:
\begin{equation}
Z^{(a)}\sim A(\omega)e^{\imath\frac{\omega}{c}r_*}+B(\omega)e^{-\imath\frac{\omega}{c}r_*},
\label{21}
\end{equation} 
with $A(\omega), B(\omega)$ generally complex constants. As far as the amplitude of (\ref{21}) remains 
sufficiently small, $h_0$ in (\ref{15}) can still remain small, but certainly this does not 
happen for $h_1$ in (\ref{14}), becoming unbounded approaching the horizon, this denoting 
a breakdown of the perturbative regime, as in \cite{21}.\\
As a consequence of the reasonings above, although solutions of (\ref{19}) could be formally 
extended up to the cosmological horizon ($r_*\rightarrow\infty $ in tortoise radial coordinate), this 
cannot hold for all perturbative functions in (\ref{3}). Summarizing, 
according to the results in \cite{21}, the perturbative regime certainly works as far as $rH<<1$. It should be also noticed that the regime $rH<<1$ does not merely correspond to 
$r_*<\infty$. In fact, from (\ref{20}), it is easy to see that the condition $rH<<1$ does imply
$\frac{\left(e^{2Hr_*}-1\right)}{\left(e^{2Hr_*}+1\right)}<<1$: this inequality is satisfied only for
$Hr_*<<1$.\\ 
A similar phenomenon can be observed in the perturbative regime of, for example, Schwarzschild
black holes near the horizon. There, the issue is solved by using Eddington-Finhelstein coordinates.
However, we stress that also by using suitable coordinates, the smallness of perturbative functions
should  be checked.\\
For all the aforementioned questions, in the next section we write down from the onset the perturbative axial equations but in terms of the tortoise coordinate $r_*$.

\section{Perturbative equations with $r_*$ coordinate}

The line element (\ref{1}) in $\{t,\phi, r_*,\theta\}$ coordinates, with $r_*$ given by (\ref{20}), becomes:
\begin{equation}
ds^2=e^{2\nu(r_*)}dt^2-e^{2\nu(r_*)}d{r^2_*}-r^2(r_*)\left[d\theta^2+\sin^2\theta d\phi^2\right].
\label{a1}
\end{equation} 
Concerning the perturbative equations, with the help of the unperturbed metric (\ref{a1}) and 
with (\ref{3}), the relevant axial equations are:
\begin{eqnarray}
& & \imath\omega e^{-2\nu(r_*)} h_{0,r_*}-2\imath\omega\frac{h_0}{r(r_*)}-
\frac{2n}{r^2(r_*)}h_1+e^{-2\nu(r_*)}\omega^2 h_1=0,\label{a2}\\
& & h_{1,r_*}-\imath\omega h_0=0. \label{a3}
\end{eqnarray} 
It is easy to show that equations (\ref{a2})-(\ref{a3}) can be managed to obtain:
\begin{eqnarray}
& & h_1(r_*.\omega)=r(r_*) Z^{(a)}(r_*,\omega), \label{a4}\\
& & h_0(r_*,\omega)=
\frac{\left(r(r_*)Z^{(a)}_{,r_*}+e^{2\nu(r_*)}Z^{(a)}\right)}{\imath\omega},\label{a5}\\
& & Z^{(a)}_{,r_*,r_*}+Z^{(a)}\left[\frac{\omega^2}{c^2}-
\frac{e^{2\nu(r_*)}}{r^2(r_*)}\ell(\ell+1)\right]=0. \label{a6}
\end{eqnarray}
Equation (\ref{a6}) is nothing else but the Regge-Wheeler one (\ref{17}).\\
We can thus study the perturbative regime for (\ref{a4})-(\ref{a6}). We must have;
\begin{eqnarray}
& & |h_1|=\left|Z^{(a)}\right|r(r_*)<<1, \label{a7}\\
& & |h_0|=\left|\frac{r(r_*)Z^{(a)}_{,r_*}+e^{2\nu(r_*)}Z^{(a)}}{\imath\omega}\right|<<1\label{a8}
\end{eqnarray}
The conditions abobe can be certainly satisfied for small values of $r_*<<1$. Since
$Sup\{r(r_*)\}=\frac{1}{H}$, from (\ref{a7}) we deduce that $\left|Z^{(a)}\right|<<H$. For $r_*\rightarrow\infty$ we have the behavior (\ref{21}). From (\ref{a8}) we realize that 
certainly for sufficiently hight frequencies $|\omega|$, i.e. for $|\omega\geq cH|$, and for initial conditions such that $Max\{|A(\omega)|,|B(\omega)|\}<<H$, where $A(\omega), B(\omega)$ are the same
quantities 
of equation (\ref{21}), the perturbative regime holds. For low frequencies modes with
$|\omega|<<cH$, thanks to the term $e^{2\nu(r_*)}Z^{(a)}$ in (\ref{a8}), although we have that
$\lim_{r_*\rightarrow\infty}e^{2\nu(r_*)}=0$, perturbative regime could be no longer available. However,
thanks to the presence of the cosmological horizon at $r=1/H$, frequencies modes with 
$|\omega|<cH$ are not practically observable. As a result, in the coordinates $\{t,\phi, r_*,\theta\}$
the perturbative regime can be extended to all observable frequencies $\omega$ provided that
$\left|Z^{(a)}\right|<<H$.\\
The conditions discussed above for the range of validity of the perturbative regime should be always done in order to assure that the equations involved can be applied to the whole range of validity of the used coordinates. In effect, also if the Regge-Wheeler solutions remain in the range of validity
of the perturbative regime, it should be checked that all perturbative functions remain sufficiently small. Otherwise, the Regge-Wheeler equation cannot longer be used to study the linear regime. This fact is often missing in the literature.

\section{A possible quantum spectrum for Regge-Wheeler equation}

As shown in section above, perturbative regime in appropriate coordinates can be fulfilled up to  cosmological horizon, provided that suitable boundary conditions are imposed.
In the following, in order to explore a new possible discrete spectrum 
denoting possible quantum features of gravitational waves at scales where
quantum fluctuations come into action, we are interested to study the (\ref{a4})-(\ref{a6}) at small distances with respect 
to the cosmological horizon by adopting the condition $Hr_*<<1$.
Within this 
perturbative regime, we can expand (\ref{a6}) and thus express $r$ with respect to $r_*$ in a series 
expansion. We obtain:
\begin{equation}
r=r_*-\frac{H^2}{3}r_*^3+\frac{2H^4}{15}r_*^5-\frac{17H^6}{315}r_*^7+o(1),\;\;
r_*\in[0,\frac{\epsilon}{H}],\;\epsilon<<1.
\label{22}
\end{equation}
Thanks to (\ref{22}), the term $\frac{\ell(\ell+1)}{r^2}$ in (\ref{a6}) or
(\ref{19}) can be taylor expanded in the following way:
\begin{eqnarray}
& &\frac{\ell(\ell+1)}{r^2}=
\frac{\ell(\ell+1)}{r_*^2{\left(1-\frac{H^2}{3}r_*^2+
\frac{2H^4}{15}r_*^4-\frac{17H^6}{315}r_*^6+o(1)\right)}^2}=\nonumber\\
& &=\frac{\ell(\ell+1)}{r_*^2}+\frac{2\ell(\ell+1)}{3}H^2+\frac{H^4\ell(\ell+1)}{15}r_*^2+o(1)
\label{23}
\end{eqnarray}
As a conequence of (\ref{23}), in the region with 
$Hr_*<<1$ the principal part of the  Regge-Wheeler equation
(\ref{19}) becomes:
\begin{eqnarray}
& &Z^{(a)}_{,r_*,r_*}+Z^{(a)}\left[\frac{\omega_{\Lambda}^2}{c^2}-
\frac{1}{r_*^2}\ell(\ell+1)-V^{(a)}(r_*)\right]=0, \label{24}\\
& &\frac{\omega_{\Lambda}^2}{c^2}=\frac{\omega^2}{c^2}+\frac{H^2}{3}\ell(\ell+1),\;\;
V^{(a)}(r_*)=\frac{H^4}{15}\ell(\ell+1)r_*^2+o(1). \label{25}
\end{eqnarray}
As a result, the effective Regge-Wheeler equation at short distances looks like a 
Schrodinger equation with the potential of a 3D harmonic oscillator in the radial 
tortoise coordinate.\\
As an interesting application of the 'effective' equation (\ref{24}), we apply to (\ref{24}) the same quantization procedure suitable for a 3D harmonic oscillator. This is justified by the fact that short distances are suitable in order to study possible quantum effects or quantum modes related to
GW by imposing boundary conditions different from the ones leading to quasinormal modes.\\
To start with, after posing
$B^4=\frac{H^4}{15}\ell(\ell+1),\frac{\omega_{\Lambda}^2}{c^2}=E_{\Lambda,\ell}$, for (\ref{24}) we  have:
\begin{equation}
Z^{(a)}_{,r_*,r_*}+Z^{(a)}\left[E_{\Lambda,\ell}-
\frac{1}{r_*^2}\ell(\ell+1)-B^4 r_*^2\right]=0. \label{26}
\end{equation}
For large $r_*$, the principle part of the solution of (\ref{26}) is dominated by 
$e^{B^2 r_*^2/2}, e^{-B^2 r_*^2/2}$. In ordinary quantum mechanics the behavior $e^{B^2 r_*^2/2}$
is discarged since the wave function must be square integrable. In our context $Z^{(a)}$ is not a wave
function. Nevertheless, in order to search possible quantum modes, decaying modes with
$e^{-B^2 r_*^2/2}$ exactly mimicking the quantum case are certainly a reasonable choice. 
With the choice above, the solution of (\ref{26}) can be obtained with the position:
\begin{equation}
Z^{(a)}=e^{-B^2 r_*^2/2} F^{(a)}(r_*). 
\label{27}
\end{equation}
From (\ref{26}) we obtain:
\begin{equation}
F^{(a)}_{,r_*,r_*}-2B^2 r_*F^{(a)}_{,r_*}+
F^{(a)}\left[E_{\Lambda,\ell}-B^2-\frac{\ell(\ell+1}{r_*^2})\right],\;
F^{(a)}(r_*=0)=0.\label{28}
\end{equation}
Equation (\ref{28}) can be solved by a series expansion in the following way:
\begin{equation}
F^{(a)}=r_*^{s}\sum_{i=0}^{\infty}a_i r_*^i,\;\;\;a_0\neq 0. \label{29}
\end{equation}
By substituing (\ref{29}) in (\ref{28}), we obtain, for the lowest order term in $r_*^{s-2}$, the condition:
\begin{equation}
\left[s(s-1)-\ell(\ell+1)\right]a_0=0\rightarrow s=(\ell+1).
\label{30}
\end{equation}
At the order $r_*^{s-1}$ we have 
$\left[s(s+1)-\ell(\ell+1)\right]a_1=0\rightarrow a_1=0$. By further proceeding we find that  
all coefficients with odd indices $i$ are vanishing, while for the ones with even indices we obtain
the recurrence formula:
\begin{equation}
\frac{a_{i+2}}{a_i}=\frac{\left[\left(2i+2\ell+3\right)B^2-E_{\Lambda,\ell}\right]}
{\left(i+2\right)\left(i+2\ell+3\right)}.
\label{31}
\end{equation} 
Hence, for $Z^{(a)}$ we obtain:
\begin{equation}
Z^{(a)}=e^{-B^2 r_*^2/2} r_*^{\ell+1}\left[a_0+a_2 r_*^2+a_4 r_*^4+\cdots\right].
\label{32} 
\end{equation}
Since
$\lim_{i\rightarrow\infty}\frac{a_{i+2}}{a_i}\sim\frac{2B^2}{i}$, one has that 
$F^{(a)}\sim e^{B^2 r_*^2}$ and consequently in the usual theory of a 3D harmonic oscillator
a further condition is imposed: series expansion in (\ref{29}) with (\ref{31}) stops end for a certain even integer index $k, k\geq 0$, assuring once again the square integrability of the wave equation. By mimicking the quantization procedure, we impose the same condition.\\
Hence, we suppose that there exists an even non negative integer $k$ such that 
\begin{equation}
E_{\Lambda,k,\ell}=\left(2k+2\ell+3\right)B^2.
\label{33}
\end{equation}
Finally, equation (\ref{33}) is nothing else but
\begin{equation}
\omega_{k,\ell}^2=c^2 H^2\left[\frac{\left(2k+2\ell+3\right)}{\sqrt{15}}
\sqrt{\ell(\ell+1)}-\frac{\ell(\ell+1)}{3}\right].
\label{34}
\end{equation}
Equation (\ref{34}), with the help of (\ref{25}), can also be written as
\begin{equation}
\frac{\omega_{\Lambda}^2}{c^2}=
\left(k+\ell+\frac{3}{2}\right)\sqrt{2V^{(a)}_{,r_*,r_*}(r_*=0)},
\label{a34}
\end{equation}
where the role of the potential $V^{(a)}$, as happens for the quasinormal modes calculated by means of the WKB approximation, (see for example \cite{10,11,12}) is highlighted.\\
First of all, note that, differently from 
quasinormal modes, the right hand side of (\ref{34}) is strictly positive for 
$k\geq 0, \ell\geq 2$ and as a consequence the frequency spectrum $\{\omega_{k,\ell}\}$ is always real.
Hence, we can write:
\begin{equation}
|\omega_{k,\ell}|=cH\sqrt{\left[\frac{\left(2k+2\ell+3\right)}{\sqrt{15}}
\sqrt{\ell(\ell+1)}-\frac{\ell(\ell+1)}{3}\right]}
\label{35}
\end{equation}
Also note that for $\ell>>k$ we have that 
$\omega_{k,\ell}\sim cH\ell\sqrt{\frac{2}{\sqrt{15}}-\frac{1}{3}}$.\\
It can also be interesting to calculate the magnitudo of the frequency level spacing
$\omega_{k,\ell}-\omega_{k, \ell-1}$ for $\ell\geq 3$ (GW case.). For the lowest order mode with $k=0$, the frequency level spacing magnitude reaches an asymptotic value for
$\ell\rightarrow\infty$ given by $cH\sqrt{\frac{2}{\sqrt{15}}-\frac{1}{3}}$. However, it is easy to see 
that this asymptotic value represents a very good approximation also for low values of $\ell$. 
Consequently, for $\ell\geq 3$ we have:
\begin{equation}
\omega_{0,\ell}-\omega_{0, \ell-1}\simeq cH\sqrt{\frac{2}{\sqrt{15}}-\frac{1}{3}},\;\;
\forall\ell\geq 3.
\label{36}
\end{equation}
For $k=0$ with $\Delta\ell=n\in N$ we thus obtain 
\begin{equation}
\omega_{0,\ell}-\omega_{0, \ell-n}\simeq cHn\sqrt{\frac{2}{\sqrt{15}}-\frac{1}{3}},
\forall\ell\geq 3,\;\ell-n\geq 2
\label{36a}
\end{equation}
Moreover, $\forall k\in\Re$, it follows that
$\lim_{\ell\rightarrow\infty}\left(\omega_{k,\ell}-\omega_{k, \ell-1}\right)=
cH\sqrt{\frac{2}{\sqrt{15}}-\frac{1}{3}}$. Concerning the frequency level spacing with respect
to $k$ given by $\Delta k=\pm 2$ we obtain:
\begin{equation}
\lim_{k\rightarrow\infty}\left(\omega_{k,\ell}-\omega_{k-2, \ell}\right)=0,\;\;
\lim_{\ell\rightarrow\infty}\left(\omega_{k,\ell}-\omega_{k-2, \ell}\right)=
\frac{cH}{\sqrt{2\sqrt{15}-5}}.\label{37}
\end{equation}
Formula (\ref{36}) does imply that for the fundamental mode with $k=0$ 
the frequency level spacing is practically independent on $\ell$.
This happens also $\forall k\in\Re$ but with $\ell>>k$ for $k\geq 1$. The first of 
(\ref{37}) means that, although $\lim_{k\rightarrow\infty}\omega_{k,\ell}=\infty$,
the frequency level spacing $\Delta k=\pm 2$
$\forall \ell\in\Re$ becomes asymptotically zero, while the second equation does imply that, 
$\forall k\in\Re$, frequency level spacing $\Delta k=\pm 2$ is practically independent on $\ell$, provided that
$\ell>>k$.\\
From the study above it emerges that the discrete spectrum (\ref{35}) is certainly different
from the quasinormal one. In particular, the spectrum (\ref{35}) is real and with a 
non linear dependence on the indices $k,\ell$. The so obtained real spectrum is not in contradiction with 
results in \cite{R1}, where normal modes with real frequencies are forbidden in a 4-dimensional
de Sitter spacetime. In fact, more generally, quasinormal modes are obtained by imposing purely 
outgoing waves at spatial infinity for the asymptotic plane waves solution given by (\ref{21}).
Conversely, we are exploring modes with possible quantum signatures due to quantum spacetime 
fluctuations. There, it is reasonable to suppose that 
aforementioned quantum modes are characterized by the condition that the Regge-Wheeler
function is decreasing
with respect to $r_*$. From a mathematical point of view, in the language of WKB approximation, we have a situation
where $Z^{(a)}(r,\omega)\sim 0$ for $r>\overline{r},\;\overline{r}<<1/H$ with 
$\lim_{r_*\rightarrow\infty} Z^{(a)}(r_*,\omega)=0$.\\
From a physical point of view,
with the actual estimated value for $\Lambda$, from (\ref{35}) we obtain for the fundamental mode
$k=0$ with $\ell=2$, the value $\omega_{0,2}\sim 10^{-18}Hz$, that is the lowest possible value for 
(\ref{35}). These ultra low frequency values are 
compatible with the supposed ones due to primordial GW (see for example 
\cite{22} and references therein) inducing the B-mode polarization of the CMBR and exptected
in the range $(10^{-18}-10^{-16})Hz$. More generally, it is expected that at short distances gravitons
composing GW can show their quantum nature in terms of quantum oscillators.

\section{Conclusions and final remarks}
 
Often in the literature, for considerations 
regarding quasinormal modes (see for example  \cite{6,7,8,9}), only the Regge-Wheeler equation is considered without analysing all perturbation functions 
depicting the perturbed metric. As argumented in this paper, we can find solutions of the Regge-Wheeler equation that are within the pertutbative regime for all values available for radial coordinate
$r$. However, this does not
happen for all functions depicting perturbations, at least with background metric given by (\ref{1}). In this regard, it is questionable that boundary conditions are imposed at
$r_*\rightarrow\infty$, in order for example to find quasinormal modes, without checking the validity of the perturbative regime. This line of research has been adopted in \cite{21}, where with the background metric (\ref{1}) the effects of a GW have been analysed up to the radial radius where perturbative regime holds.\\
We have thus derived perturbative equations in terns of the tortoise coordinate $r_*$ from the onset and we shown that perturbative regime can be valid up to the cosmological horizon, provided that suitable 
boundary conditions are chosen. As an application, in order to explore possible quantum effects for GW in a de Sitter spacetime, we obtain an effective Regge-Wheeler equation, namely equation (\ref{24}),
suitable at short distances with $Hr_*<<1$.
We obtained a discrete spectrum for (\ref{25}) by adopting the same procedure
leading to the quantization of a 3D harmonic oscillator. The outcome is provided  by the discrete spectrum (\ref{35}). The resulting spectrum is discrete and different from the quasinormal ones obtained by imposing a purely outgoing wave for $r_*\rightarrow\infty$. Moreover, the modes so obtained in this paper are characterized by a vanishing of $Z^{(a)}(r,\omega)$ for $r_*\rightarrow\infty$.
We may also suggest another genesis for the aforementioned quantum modes. In fact, we may suppose the existence of a quantum scale where the effective Regge-Wheeler equation (\ref{26}) holds with a real spectrum given by (\ref{35}). The crossover to classicality may be depicted as a solution of 
Regge-Wheeler equation where the spectrum (\ref{35}) becomes purely immaginary, i.e.
$\omega_{k,\ell}\rightarrow \imath\;\omega_{k,\ell}$, in such a way that for 
$r_*\rightarrow\infty$ we have that $Z^{(a)}(r,\omega)\sim e^{-|\omega_{k,\ell}|r_*/c}$. This suggestion
can be matter for further investigations.\\
Moreover, we have studied some properties of the so obtained spectrum, in particular the ones related to the frequency level spacing with $\Delta\ell=\pm 1$ with $k$ fixed and $\Delta k=\pm 2$
with $\ell$ fixed together with the limit for $\ell\rightarrow\infty$. It results that the frequency level spacing
for $k=0$ (fundamental mode) is practically independent on $\ell$.\\
Also note that we could infer the value of $\Lambda$ by measuring the frequency level spacing given by
(\ref{36}) for the fundamental mode of the spectrum (\ref{35}).
Given the actual value for
$\Lambda$, we speculated on the possible link of
(\ref{35}) with the spectrum provided by primordial GW inducing the B-mode
polarization of the CMBR or more generally on a possible manifestation of quantum effects
for GW at short distances. In practice, gravitons at sufficiently small distances show their 
quantum nature behaving as quantum harmonic oscillators.

\end{document}